\documentclass[manuscript]{aastex}

\usepackage{lineno}
\usepackage{url}
\linenumbers

\newcommand{\psim}{\lower.5ex\hbox{$\; \buildrel \propto \over\sim \;$}}
\newcommand{\lbar}{\lower.0ex\hbox{$\; \buildrel
{\lower0.0ex \hbox{-}} \over\lambda  \;$}}

\slugcomment{To appear in the Astrophysical Journal}

\shorttitle{Multi-Wavelength Observations of PKS\,2142$-$75}
\shortauthors{Dutka, Ojha, Pottschmidt et al.}

\begin{document}


\title{Multi-Wavelength Observations of PKS\,2142$-$75 during Active and Quiescent Gamma-Ray States}


\author{Michael S.\ Dutka}
\affil{The Catholic University of America, 620 Michigan Ave., N.E.  Washington, DC 20064}
\email{ditko86@gmail.com}

\author{Roopesh Ojha\altaffilmark{1,2}}
\affil{ORAU/NASA Goddard Space Flight Center, Astrophysics Science Division, Code 661, Greenbelt, MD 20771}

\author{Katja Pottschmidt}
\affil{Center for Research and Exploration in Space Science and Technology (CRESST), University of Maryland Baltimore Campus (UMBC) and NASA Goddard 
Space Flight Center, Astrophysics Science Division, Code 661, Greenbelt, MD 20771}

\author{Justin D.\ Finke}
\affil{Naval Research Laboratory, Space Science Division, Code 7653, 4555 Overlook Ave. SW, Washington, DC 20375}

\author{Jamie Stevens}
\affil{CSIRO Astronomy and Space Science, Locked Bag 194, Narrabri NSW 2390, Australia}

\author{Philip G.\ Edwards}
\affil{CSIRO Astronomy and Space Science, PO Box 76, Epping NSW 1710, Australia}

\author{Jay Blanchard}
\affil{Departamento de Astronom'a Universidad de Concepci—n, Casilla 160 C, Concepci\'{o}n, Chile}

\author{James E. J. Lovell}
\affil{University of Tasmania 
	School of Mathematics \& Physics
	Private Bag 37
Hobart Tas 7001, Australia}

\author{Roberto Nesci}
\affil{INAF/IAPS, via Fosso del Cavaliere 100, I 00133 Roma Italy}

\author{Matthias Kadler}
\affil{Lehrstuhl f\"{u}r Astronomie, Universit\"{a}t W\"{u}rzburg, Emil
-Fischer-Stra{\ss}e 31, D-97074 W\"urzburg, Germany}

\author{Joern Wilms}
\affil{Remeis Observatory \& ECAP, Sternwartstr. 7, 96049 Bamberg, Germany}

\author{Gino Tosti}
\affil{University of Perugia, Piazza Universit$\mathrm{\grave{a}}$ 1, 06123 Perugia, Italy}

\and

\author{Tapio Pursimo}
\affil{Nordic Optical Telescope, Apartado 474, E-38700 Santa Cruz de La Palma Santa Cruz de Tenerife, Spain}

\author{Felicia Krauss}
\affil{Remeis Observatory \& ECAP, Sternwartstr. 7, 96049 Bamberg, Germany}

\author{Cornelia M\"uller}
\affil{Remeis Observatory \& ECAP, Sternwartstr. 7, 96049 Bamberg, Germany and  Lehrstuhl f\"{u}r Astronomie, Universit\"{a}t W\"{u}rzburg, Emil
-Fischer-Stra{\ss}e 31, D-97074 W\"urzburg, Germany}

\author{Neil Gehrels}
\affil{NASA Goddard Space Flight Center, Astrophysics Science Division, Code 661, Greenbelt, MD 20771}


\altaffiltext{1}{Adjunct Professor, The Catholic University of America, 620 Michigan Ave., N.E.  Washington, DC 20064}
\altaffiltext{2}{Honorary Fellow, University of Tasmania, Newnham Dr, Newnham TAS 7248, Australia}


\begin{abstract}
PKS\,2142$-$75 (a.k.a. 2FGL \,J2147.4$-$7534) is a flat-spectrum
radio quasar that was observed quasi-simultaneously by a suite of
instruments across the electromagnetic spectrum during two flaring 
states in April 2010 and August 2011 as well as a quiescent state
in December 2011 through January 2012.  The results of these campaigns and 
model spectral energy distributions (SEDs) from the active and quiescent
states are presented.  The SED model parameters of PKS\,2142$-$75 indicate that the two
flares of the source are created by unique physical conditions.  SED
studies of flat spectrum radio quasars are beginning to indicate that
there might be two types of flares, those that can be described
purely by changes in the electron distribution and those that require
changes in other parameters, such as the magnetic field strength or
the size of the emitting region.
\end{abstract}


\keywords{galaxies: active --- quasars: individual (PKS\,2142$-$75) --- gamma rays: observations --- 
X-rays: individual (PKS\,2142$-$75) --- ultraviolet: galaxies --- radio continuum: galaxies}

\section{Introduction}\label{introduction}

Active galactic nuclei (AGN) are the brightest persistent sources of
electromagnetic radiation in the Universe. They radiate intensely from
low-energy radio waves to very-high-energy $\gamma$-rays and are known to
be highly variable at all frequencies.  Blazars are the most luminous
and violently variable subclass of AGN. High and highly variable
polarization has been detected in the radio and optical emission of
blazars, and they are the most common type of AGN detected at GeV
energies \citep{Abdo2010a}.  Blazars have extended plasma jets,
which can be detected at radio and, in a growing number of cases, at
optical and X-ray frequencies as well\footnote{Lists of X-ray detected
and optically-detected jets can be found at
\url{http://hea-www.harvard.edu/XJET/} and
\url{http://astro.fit.edu/jets/}, respectively.}. Observational data
suggest that blazars are those radio-loud AGN whose jets are aligned
at small angles with respect to the line of sight.  In this scenario,
the bulk relativistic motion of the emitting material results in
``relativistic beaming", i.e. both the luminosity and the variability
of emission appear to be enhanced compared to what would be measured
in the rest frame of the blazar \citep[e.g.,][]{Urry1995}.  The jets
can show apparent superluminal motion in high resolution very long
baseline interferometric (VLBI) radio images, a consequence of the
jets pointing at a small angle to the line of sight. Blazars can be
subdivided, based on their spectral features, into BL\,Lacertae (BL
Lac) objects and flat spectrum radio quasars (FSRQs).  BL\,Lac objects
exhibit weak or no emission lines, while FSRQs show strong, broad
emission lines \citep{Urry1995}.

The SEDs of blazars show a characteristic ``double-bumped" structure,
with the lower frequency bump almost certainly originating from beamed
electron synchrotron emission from the jet \citep{Urry1995}.  Aside
from their classification based on their optical spectra, blazars are
also classified based on their frequencies of peak synchrotron
emission in a $\nu F_{\nu}$ representation.  Low-synchrotron peaked
(LSP) blazars have $\nu^{\textnormal{\small sync}}_{\textnormal{\small
peak}}$ at $<10^{14}$\ Hz; high-synchrotron peaked (HSP) blazars have
$\nu^{\textnormal{\small sync}}_{\textnormal{\small peak}}$
$>10^{15}$\ Hz; and intermediate-synchrotron peaked (ISP) blazars have
$\nu^{\textnormal{\small sync}}_{\textnormal{\small peak}}$ between
$10^{14}$ and $10^{15}$\ Hz \citep{Abdo2010_sed}.  BL\,Lac objects can
be any of the three classes; however almost all FSRQs are LSP blazars.

The origin of the higher frequency bump, peaking in $\gamma$ rays, is
less well understood.  Two families of models attempt
to describe them: leptonic and hadronic models
\citep{Bottcher2007}. Leptonic models invoke a source of soft seed
photons being Compton up-scattered to higher energies by relativistic
electrons within the jet.  The seed photon source could be synchrotron
emission from the same electrons which scatter them (known as
synchrotron self-Compton or SSC), or could be external
to the jet (external Compton or EC).  The accretion disk
\citep{Dermer92}, the broad line region \citep[BLR;
e.g.,][]{Sikora94}, the dust torus \citep{Blazejowski2000}, or even
the cosmic microwave background \citep{Bottcher2008} have all been
considered as seed photon sources for EC processes.  Hadronic models
assume that a significant fraction of the jet power is converted into
the acceleration of protons, so that the protons reach the threshold
for pion production.  The pions and their secondaries can thus create
the $\gamma$-ray emission \citep{Mannheim1992}.  Protons at these high
energies can also emit $\gamma$ rays through synchrotron emission
\citep[e.g.,][]{Muecke2003}.

Simultaneous flux density measurements of blazars at multiple
wavelengths across the electromagnetic spectrum can allow us to
distinguish among the many different high-energy emission models
proposed \citep{Bottcher2007}.  The launch of the {\em Fermi Gamma Ray
Space Telescope} \citep{Atwood2009} has created new opportunities for
studying blazars at high energies.  The {\em Fermi}/LAT (Large Area
Telescope) is more sensitive and has better energy and spatial
resolution than any previous telescope in the MeV--GeV energy range
\citep{Atwood2009}.  The LAT has put better constraints on blazar
variability within its observing window and has shown that it is not
unusual for blazars to exhibit fast variability on timescales of days
or even hours at GeV energies \citep[e.g.,][]{Saito2013}.  The LAT
observes the entire sky every three hours, which makes it an ideal
instrument on which to support multi-instrument observing campaigns on
sources in response to their changing $\gamma$-ray state.

PKS\,2142$-$75 is an FSRQ at a redshift of z$=$1.139\footnote{At this redshift, using a
cosmology with $H_0=71$\ km s$^{-1}$\ Mpc$^{-3}$, $\Omega_m=0.27$, and
$\Omega_\Lambda=0.73$, PKS\,2142$-$75 has a luminosity distance of
$d_L=7.7$\ Gpc.  We use this $d_L$ for the source in this paper.}
\citep{Jauncey1978}.  It is in the Second {\em Fermi} LAT Catalog as 2FGL
J\,2147.4$-$7534 \citep{Nolan2012} with a flux of
$3.6\pm0.2\times10^{-9}$\ ph cm$^{-2}$\ s$^{-1}$ in the 100\,MeV to
100\,GeV range, although it is not found in the First {\em Fermi} LAT 
Catalog \citep{Abdo2010_1fgl}.  

The LAT first detected a flare from PKS\,2142$-$75 on 2010 April 4 
(MJD 55290), when the source reached a daily averaged flux of
 $(1.1 \pm 0.3) \times 10^{-6} $ph cm$^{-2}$ s$^{-1}$ \citep{Wallace2010}
 in the 100\,MeV to 300\,GeV range. This flux represented an almost 3
order of magnitude increase over the average flux reported in
the 2FGL catalog \citep{Nolan2012}. After the initial flare, this source was detected in a flaring state by the LAT in
2010 October to November and again in 2011 July to August.

%Typical $\gamma$-ray flares of
%this source reach a flux of around $1.0 \times 10^{-6}$\ ph cm$^{-2}$\
%s$^{-1}$ in the 100\,MeV--300\,GeV energy range and they happen
%infrequently.  

During the latest flaring period in 2011, a multi-wavelength observing
campaign was carried out using the Ceduna radio telescope, the
Australia Telescope Compact Array (ATCA), {\em Swift}, the Rapid Eye
Mount Telescope (REM), and LAT. These quasi-simultaneous
data were complemented with simultaneous data from the ongoing SMARTS
blazar monitoring program \citep{Bonning2012}.  The data were used to
construct a broadband SED of this object in its active state.  We
organized another quasi-simultaneous multi-wavelength campaign using
the same instruments (except for the REM) during a quiescent
$\gamma$-ray state in 2011 December - 2012 January. This is a
particular strength of this work, as quiescent SEDs are often
constructed with non-simultaneous data, which can compromise their
utility.  The simultaneous quiescent SED allowed us to determine which
components of the SED were at elevated levels during the
active state and provided insight into the dominant physical processes
occurring at different wavelengths.  We present these observations and
the resulting SEDs from the active and quiescent states supplemented
by a broadband SED constructed using archival data coincident with an
earlier flare in 2010 April.  All three SEDs were fit with physical
models and the results and their implications for high-energy blazar
emission are discussed.  Throughout this paper we will refer to the
2010 April flare as flare A and the 2011 July-August flare as flare
B.  The luminosity of this source during flare A is $3.2 \times 10^{49}$ ergs s$^{-1}$ 
in the 100\,MeV to 300\,GeV energy range, which is comparable to the luminosity of some 
of the brightest flares detected by Fermi \citep{Nalewajko2013}.  During the quiescent period 
the 100\,MeV to 300\,GeV luminosity is $4.4 \times 10^{48}$ ergs s$^{-1}$, which places 
PKS\,2142$-$75 among the most luminous 2FGL FSRQs \citep{Nolan2012}.    

%We first discuss the gamma-ray properties in section 2 and then
%proceed in ascending wavelength order.  We then discuss the spectral
%energy distribution model used for this work in section 8 which we
%follow with a discussion of the results the main conclusions.

%PKS 2142-758 coordinates
%326.8030429 -75.6036736 21h47m12.7303s -75d36m13.225s
%alternative names (looks like it was part of a number of different surveys)
%PKS 2142-75
%[HB89] 2142-758
%PMN J2147-7536
%BZQ J2147-7536
%MRC 2142-758
%SUMSS J214713-753611
%AT20G J214712-753613
%CRATES  J2147-7536
%ICRF J214712.7-753613
%IERS B2142-758
%KWP81 2142-75
%VCV2001 J214712.6-753611
%HRT J214712-753621
%MGL 2009 2541

\section{LAT Observations}\label{gamma}
The LAT, on the {\em Fermi} spacecraft is a high-energy $\gamma$-ray detector with a wide field of view ($\sim 20$\% of
the sky) covering the energy band from 20\,MeV to greater than
300\,GeV.  It primarily operates in an all sky scanning mode,
observing the entire sky every three hours.  All LAT results presented here 
were obtained using the {\em Fermi} science tools (version 09-27-00) and the
P7SOURCE\_V6 instrument response functions (IRFs).

%\begin{figure}[t]
%\includegraphics[width=90mm]{1FGLJ2152.4-7532_20bins_EsqdNdE}
%\caption{LAT energy spectrum, the model displayed is the power law fit for this source across the entire 100\,MeV to %300\,GeV energy range} 
%\end{figure}

\begin{figure}
%\epsscale{1.0} \plotone{2142-75_lightcurves}
\includegraphics[trim = 10mm 0mm 10mm 10mm, clip, width=1.0\textwidth]{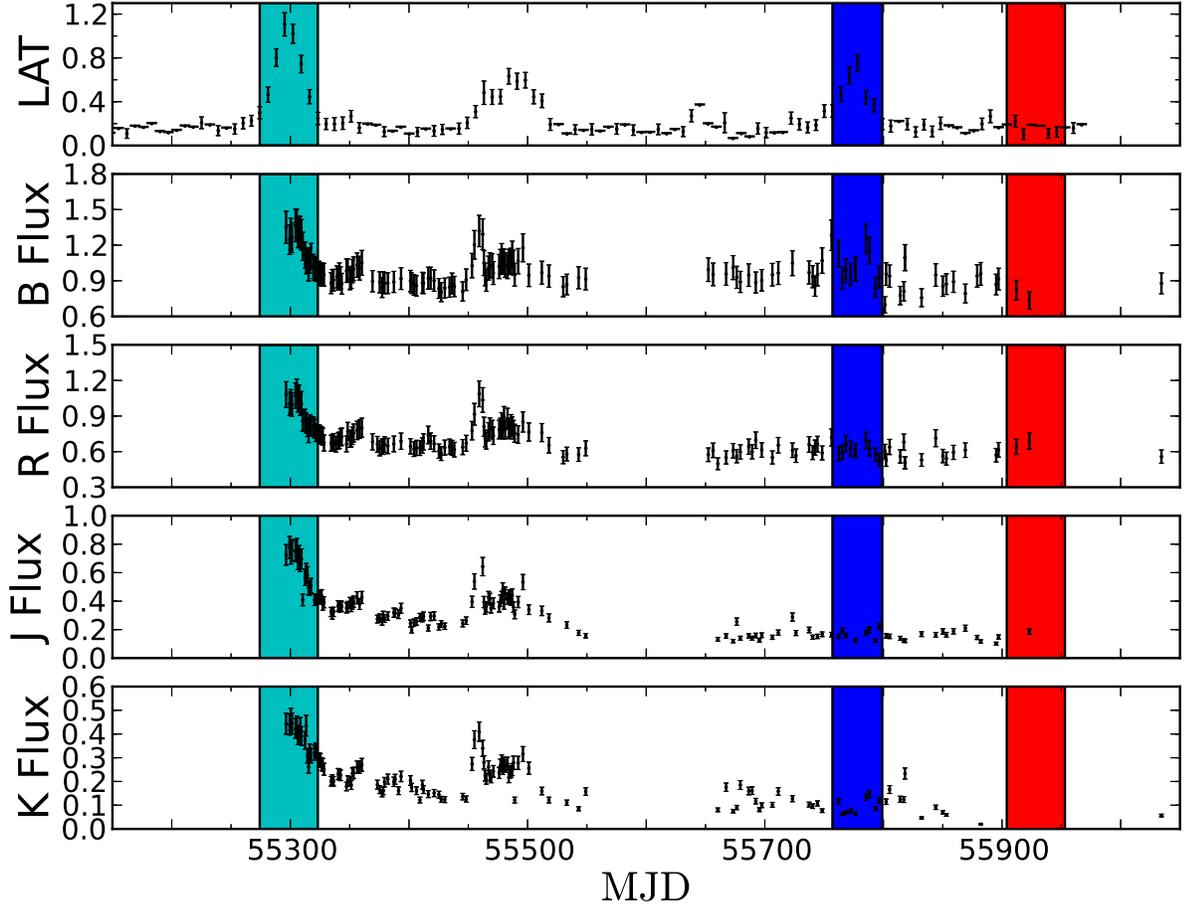}
\caption{{\em Fermi}/LAT and SMARTS light curves for PKS\,2142$-$75.  The LAT
light curve covers the 100\,MeV to 300\,GeV energy range and was generated using a bin size of 1 week and a TS threshold of 10 for upper limits, with fluxes given in units of 10$^{-6}$\ ph$^{-1}$\ cm$^{-2}$ sec$^{-1}$.  The optical light curves are from the SMARTS 1.3\,m telescope and are in units of 10$^{-15}$\ erg$^{-1}$\ cm$^{-2}$ sec$^{-1}$ \AA$^{-1}$. The
periods corresponding to the multi-wavelength observations are indicated in cyan (flare A), blue (flare B) and red (quiescent). 
\label{gammalightcurve}}
\end{figure}

\begin{figure}
\epsscale{.70}
\plotone{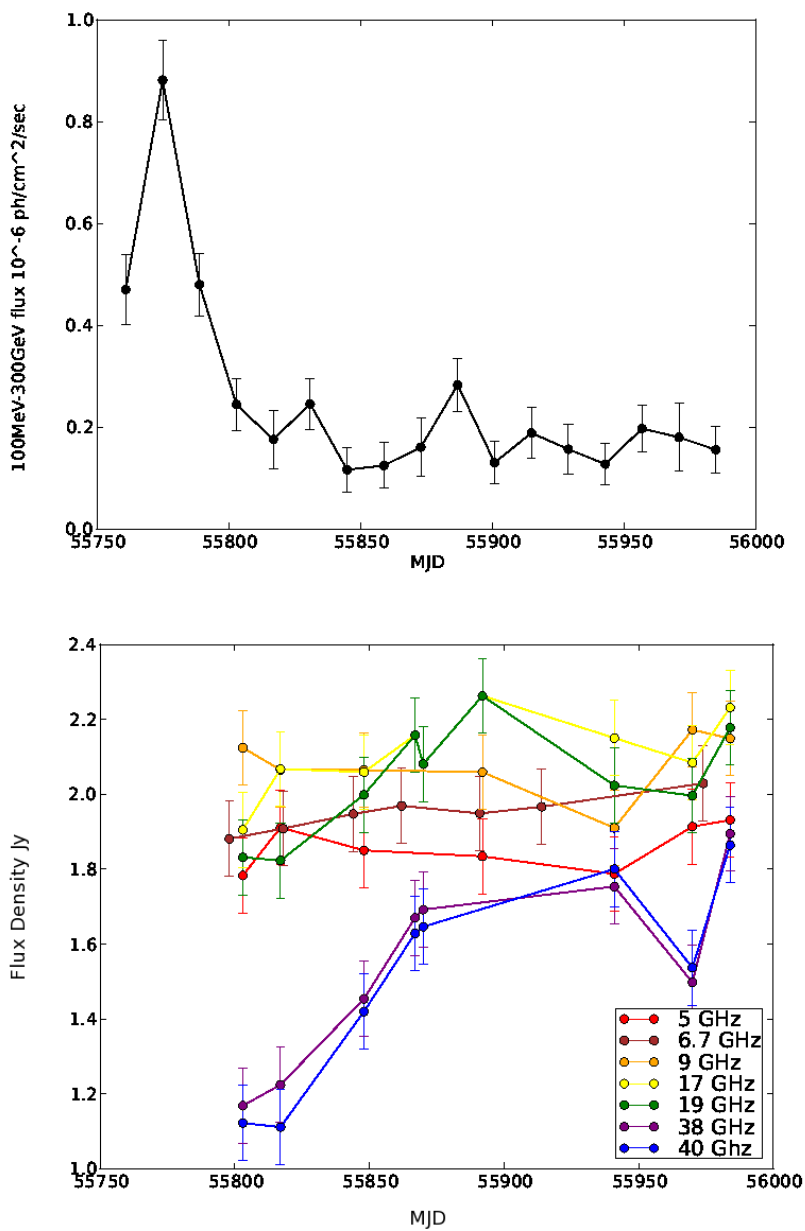}
\caption{$\gamma$-ray (above) and radio (below) light curves of PKS\,2142$-$75 displayed together for comparison purposes.  The $\gamma$-ray curve is made using 2-week bins starting on MJD 55761 (2011 July 19), which corresponds to the beginning of flare B and extends until MJD 55999 (2012 March 13). The flux densities at 6.7\,GHz were obtained from the Ceduna Radio Telescope. Those at all other frequencies were observed by the ATCA. One $\sigma$ error bars are shown. Note the significant increase in flux at the two highest radio frequencies in November 2011 following the $\gamma$-ray flare in August 2011.   \label{radiofluxes}}
\end{figure}

The light curve in the top panel of Figure.~\ref{gammalightcurve} shows
the $\gamma$-ray flux of PKS\,2142$-$75 in the 100\,MeV to 300\,GeV
energy range averaged into weekly time bins; it was created as
follows.  A circular region of interest (ROI) of radius $10^\circ$ and
``source" class events were selected for event reconstruction.  Events
above a zenith angle of $100^{\circ}$ were removed in order to
minimize the contamination from photons produced by cosmic rays interacting
with the Earth's atmosphere.  A rocking angle cut of $52^{\circ}$ was
applied. The spectral index of the source was fixed to $2.52\pm0.04$, the
value computed in the 2FGL catalog \citep{Nolan2012}.  Fitting of
the spectral index with the low photon statistics present in 1 week of
data proved to be unreliable.  Sources within the 10$\degr$ ROI in the XML model had fixed
indices and free normalizations and sources that were outside the ROI
had all spectral parameters fixed.  Sources in the 2FGL catalog were used to define the model.  
The background model includes a Galactic diffuse emission component and an isotropic component, which
are provided by the publicly available files
gll\_iem\_v02\_P7SOURCE.fit and isotropic
iem\_v02\_P7SOURCE.txt\footnote{http://fermi.gsfc.nasa.gov/ssc/data/access/lat/BackgroundModels.html}. The
isotropic component accounts for the extragalactic diffuse emission
and the residual charged-particle background.  A power law correction to the Galactic diffuse 
background was not applied and the normalizations of the isotropic and Galactic diffuse backgrounds were fixed to one.  This method yielded reasonable residuals for the fits in each individual time bin.  The light curve begins on 2008 June 26 (MJD 54643) and ends on 2012 February 14 (MJD 55971).  Upper limits on the source's flux were calculated when the test statistic
\citep[TS;][]{Mattox1996} for a given week was found to be less than 10.  The light curve
in Figure \ref{radiofluxes} was made using the same cuts on the data and
the same model for the sources in and near the ROI, although the binning 
was changed to 2 weeks. The upper limit threshold is the same and the 
time range plotted is 2011 July 19 (MJD 55761) to 2012 March 13 (MJD 55999).

\begin{figure}
\includegraphics[width=0.50\textwidth]{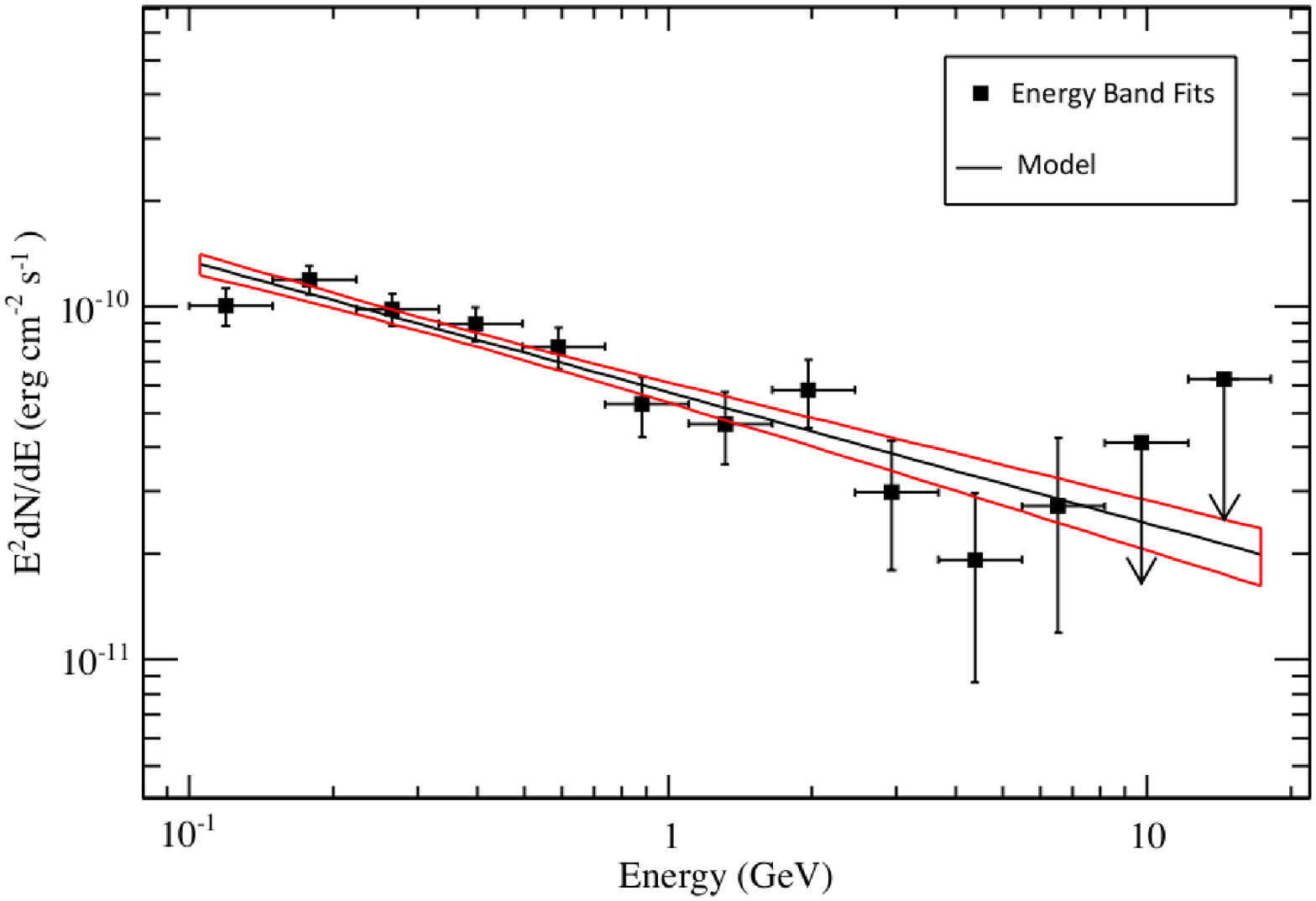}\hfill
\includegraphics[width=0.50\textwidth]{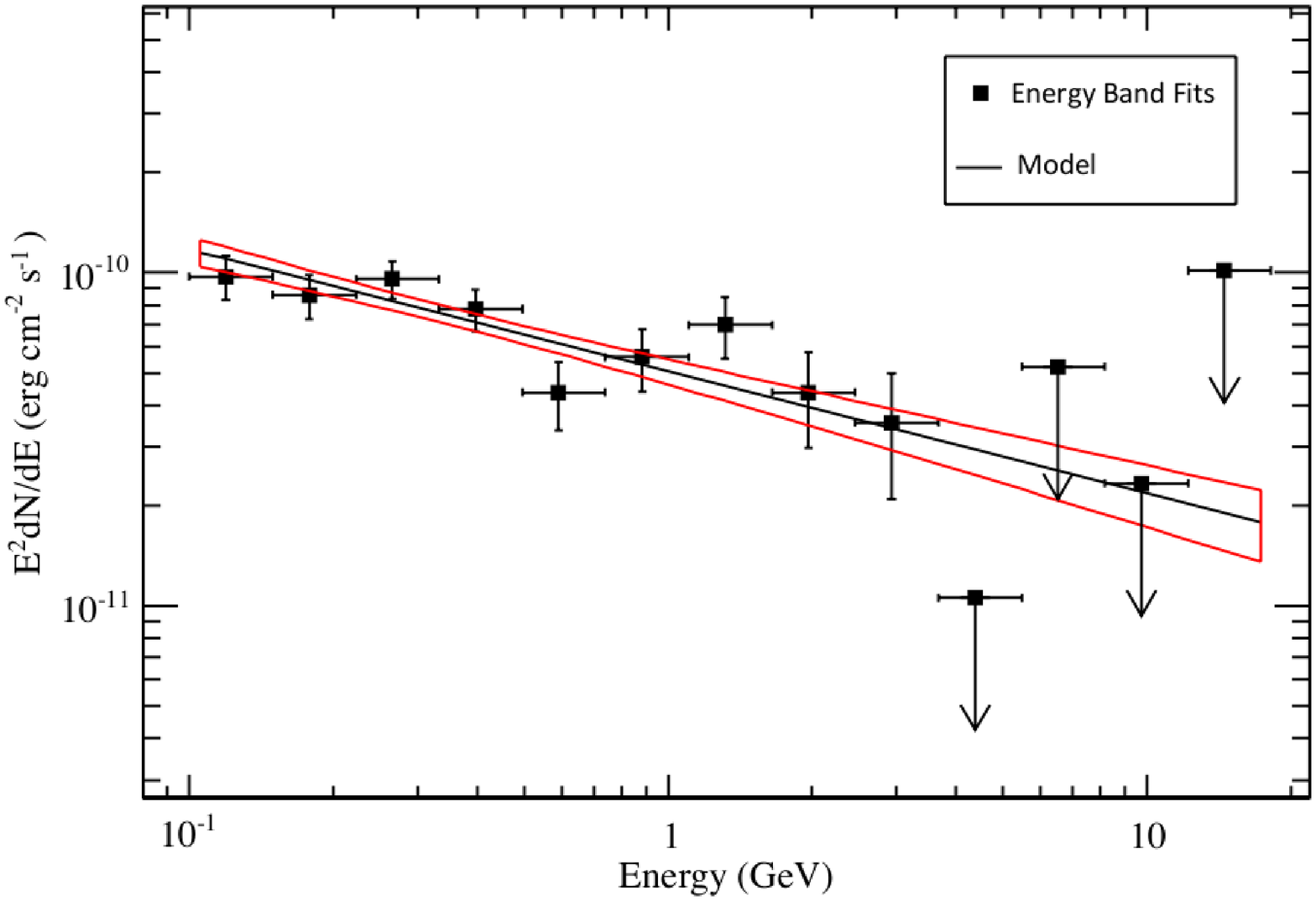}\hfill
\includegraphics[width=0.50\textwidth]{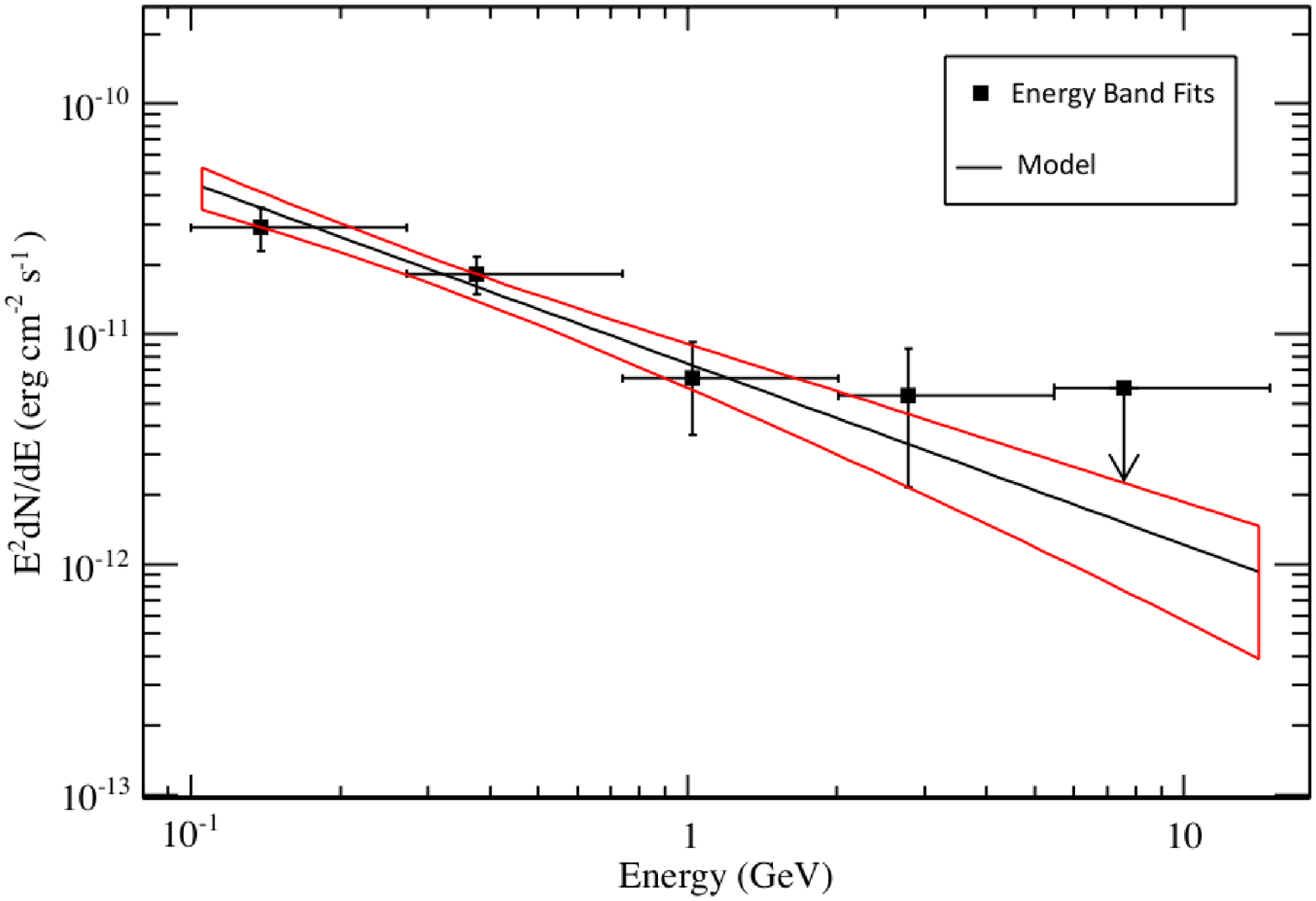}
\caption{$\gamma$-ray spectra for PKS\,2142$-$75, top left flare A, top right flare B, bottom left quiescent state.  The black curve is the initial power law fit to the full energy range and the red curve is the butterfly plot derived from the covariance values determined in the initial fit to the full energy range.
\label{gammaspectra}}
\end{figure}

During flare A the source was observed by the {\em Wide Field Infrared Survey 
Explorer} (WISE) and SMARTS.  We organized our first multi-wavelength 
campaign during flare B. Our second set of observations was
carried out in 2011 December-2012 January in order to
observe the source in a non-flaring state and obtain the quiescent
SED.  A LAT spectrum was created for each 
of these three periods by averaging and using the same cuts on the data and the same initial
model parameters used for the light curve.  The time ranges extending
from 2010 March 10 (MJD 55265) to 2010 April 30 (MJD 55316) and 2011
July 21 (MJD 55763) to 2011 August 18 (MJD 55791) were chosen to
define the flare A and flare B states because the source's daily TS
was above 25. The quiescent state was defined to be 2011 December 08
(MJD 55903) to 2012 January 26 (MJD 55952) because the source was not significantly detected in the 100\,MeV to 300\,GeV energy range on daily time scales, and this time
range coincides with our observations at other wavelengths.  For all
time ranges an initial likelihood analysis was run over the full
energy spectrum to evaluate the parameters for a power-law spectral model 
for the $\gamma$-ray emission.
For the initial likelihood analysis the spectral index was left free so
we could determine whether the spectrum changed during the
different $\gamma$-ray states.  The spectral index of the source in
the 100\,MeV to 300\,GeV energy range was found to be $2.37 \pm 0.03 $
during flare A, $2.37 \pm 0.06$ during flare B and $2.79 \pm 0.14$
during the quiescent state.  Using the initial power law as a starting
point, likelihood analysis was run again within each individual
logarithmic energy bin fixing the spectral index to the value determined from
the initial fit to the whole energy range.  The bin sizes are
different in the active state and quiescent state because
significantly fewer photons are available during the quiescent state,
which also limited the number of energy bins that could be computed.
Upper limits on the target source's flux were calculated when the TS
within the energy bin was below 10.  The $\gamma$-ray spectra are
shown in Figure \ref{gammaspectra}.

\begin{figure}
\epsscale{.99}
\plotone{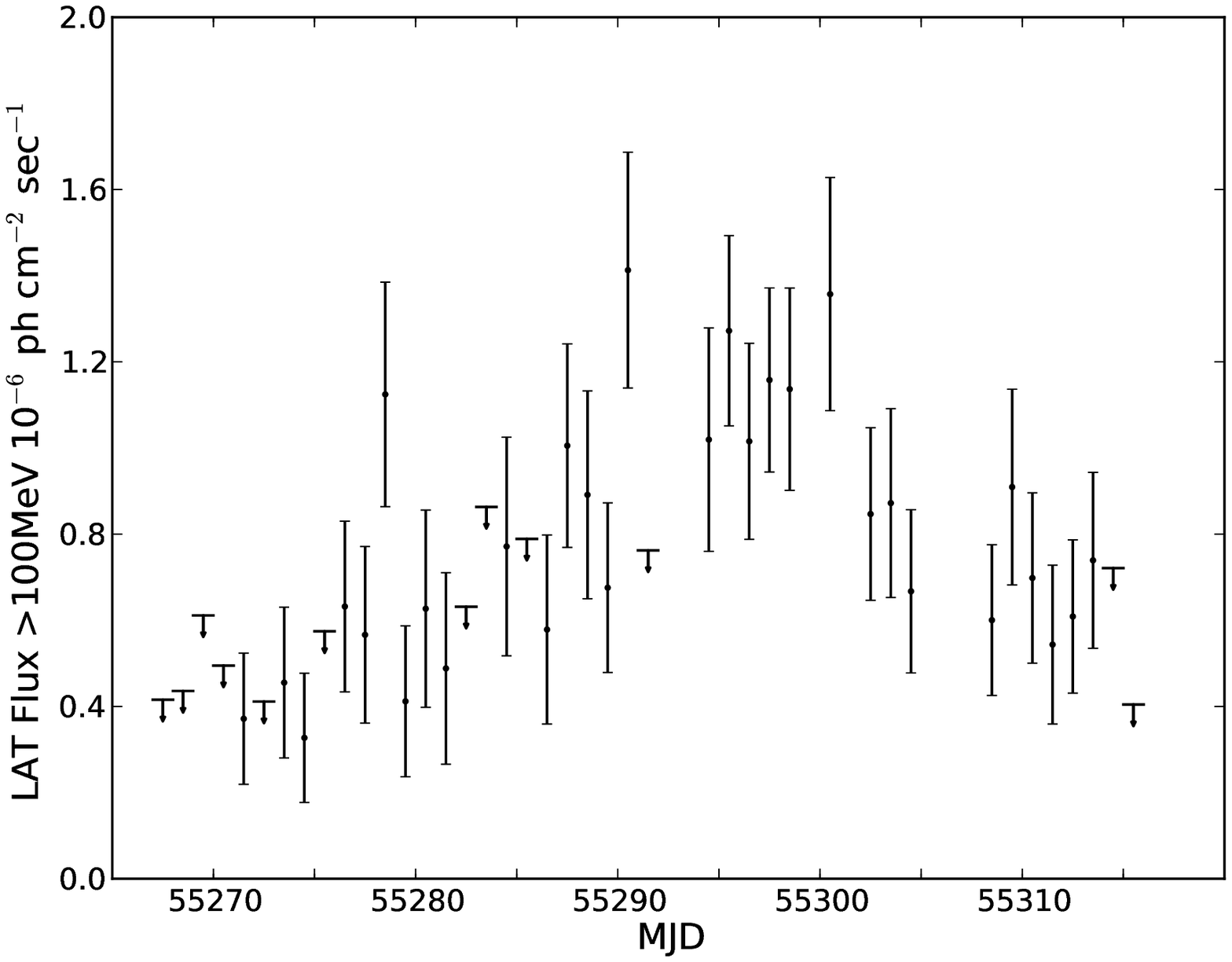}
\caption{$\gamma$-ray light curve in the 100\,MeV to 300\,GeV energy range with 1 day time bins.  This light curves starts on 2010 March 10 (MJD 55265) and ends on 2010 April 30 (MJD 55316).  The same initial cuts on the data, TS threshold for upper limits, and initial model parameters as the weekly binned light curve (Figure \ref{gammalightcurve}) were applied.  
\label{daybin}}
\end{figure}

In order to estimate the variability timescale of PKS 2142$-$75, a light curve with 1 day bins for flare A was generated and can be found in Figure \ref{daybin}.  On 2010 April 3-4 (MJD 55289-55290) the flux appears to double.  Thus a variability timescale of the order of a day is justified.  A similar variability timescale for flare B was found on 2011 July 29-30 (MJD 55771-55772).

\section{{\em Swift} Observations}\label{swift}

{\em Swift} is a multi-wavelength space-based observatory \citep{Gehrels2004} that has
three instruments: the Burst Alert Telescope (BAT), the UltraViolet and 
Optical Telescope (UVOT) and the X-Ray Telescope (XRT).
{\em Swift} uses momentum wheels to rapidly change the direction in
which it is pointing.  The observatory's primary
objective is to observe $\gamma$-ray bursts, making its ability to slew
rapidly a necessity; however this ability has become extremely useful
for studying blazars due to their rapid variability.

Two observing campaigns with the UVOT and XRT instruments on board
{\em Swift} were conducted on this object (which was too faint
to be detected by BAT). The first series of observations ran from 2011
August 10 (MJD 55783) through 2011 August 23 (MJD 55796; flare B)
while the second series ran from 2011 December 08 (MJD 55903) through
2011 December 21 (MJD 55916; quiescent state). {\em Swift} did not
observe the source during flare A.  The latest versions of the {\em
Swift} data reduction software (07Jun2011\_V6.11) and calibration data
(caldb.indx20110725) were used in our analysis.

\begin{figure}
\plottwo{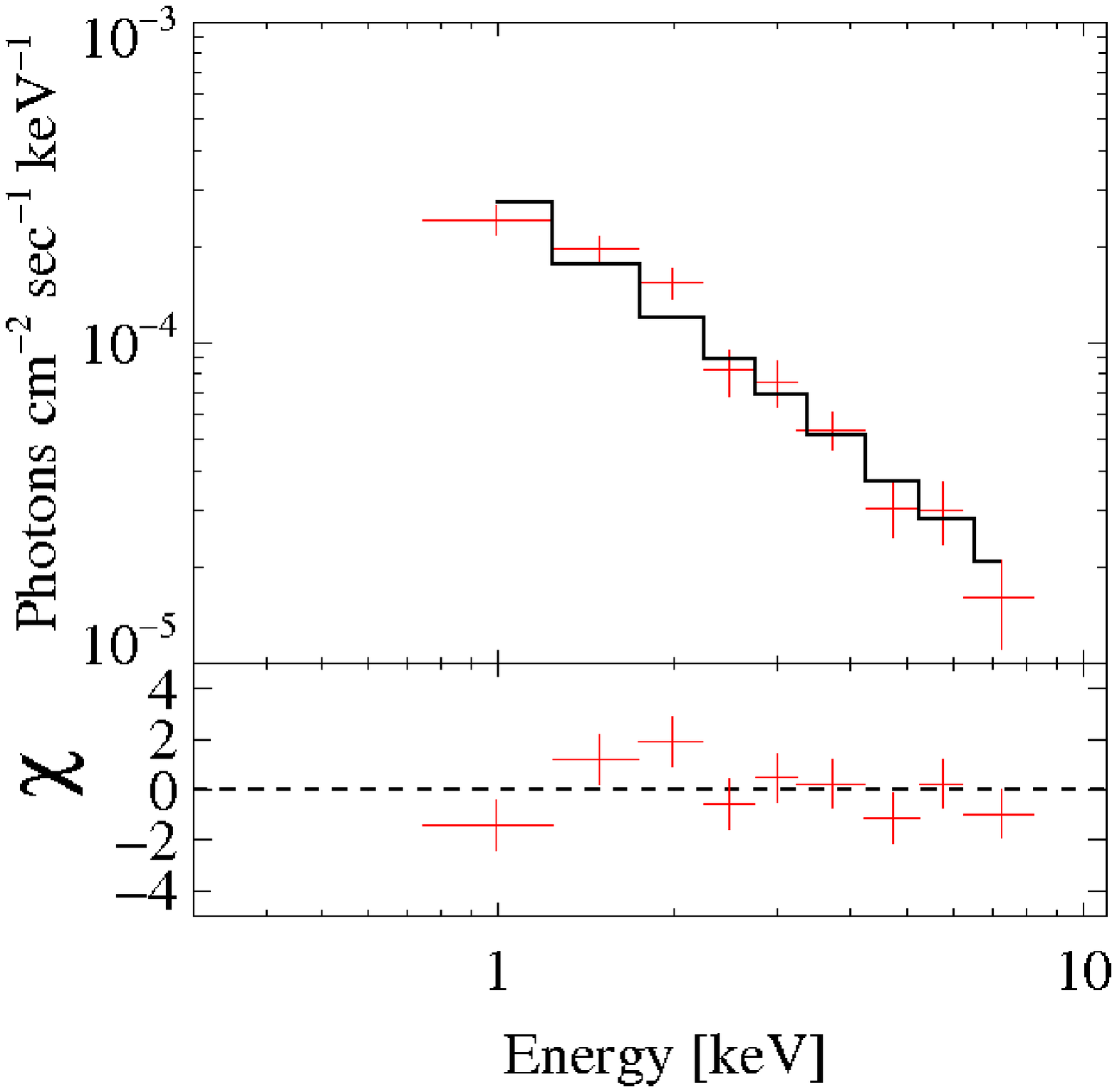}{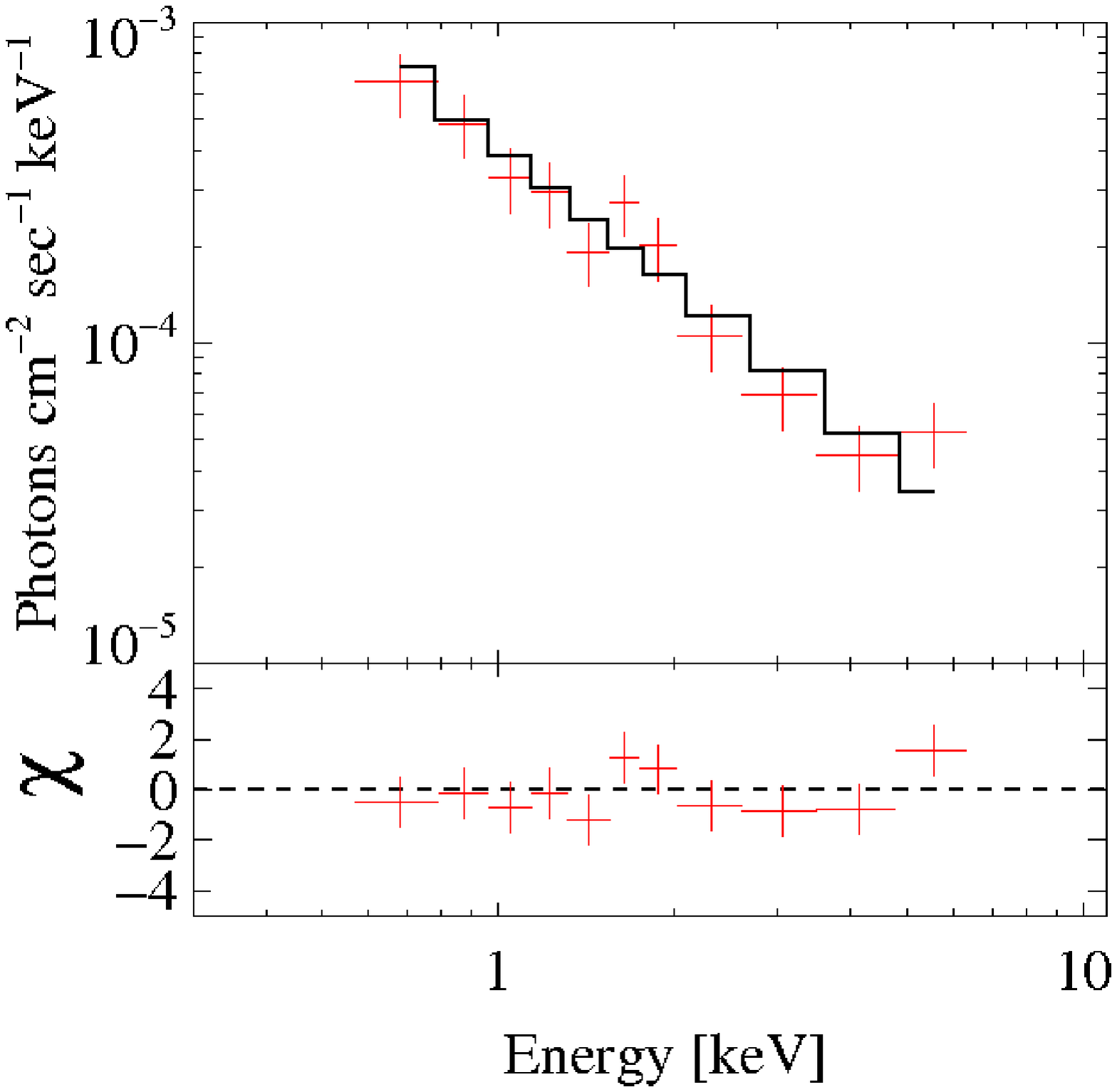}
\caption{XRT spectra for PKS\,2142$-$75. Left: flare B. Right: Quiescent.
Both states are fit with an absorbed power law of N$_{H}$\ column density of
$6.6\times10^{20}$\ cm$^{-2}$. The red points with error bars are the data, the 
black line is the fitted model, and the bottom chart shows the residuals 
in terms of $\sigma$, with 1-$\sigma$ error bars.\label{xrtfits}}
\end{figure}

Ultraviolet and optical data were taken in the `filter of the day'
mode. During the first series of observations, the source was imaged
with the $U$, $W1$ and $W2$ filters with central wavelengths of 3465
\AA, 2600 \AA\ and 1928 \AA\ respectively. During the second, the
$U$, $W1$ and $M2$ (center wavelength: 2246 \AA) filters were used.
Aperture photometry and coincidence loss corrections were made using
the standard UVOT data reduction task UVOTSOURCE.  Circular source and
background regions of radius 10\arcsec\ were used for the calibration.
Magnitudes have been corrected for Galactic extinction using the
method described in \cite{Fitzpatrick1999}.  The extinction parameter
$E(B-V)$ was calculated from the $N_{H}$\ column density using the
method described in \cite{Predehl1995}.  Uncertainties in the UV
fluxes are predominantly determined by the de-reddening correction.
For the $V$, $B$ and $U$ filters we estimate a fairly conservative
uncertainty of $5\%$\ in final calculation of the flux.  For the $W1$,
$M2$ and $W2$ filters a higher uncertainty of $25\%$\ is used because
they overlap with a UV ``bump" in the extinction curve, which is highly
dependent on the line of sight and is therefore difficult to model.

The XRT exposures from each series of observations were binned
together resulting in total exposure times of 12\,ks during flare B
and 5\,ks during the quiescent state (Figure~\ref{xrtfits}).  Circular
source and background extraction regions of radii 60\arcsec\ and
119\arcsec\ were used for all pointings.  For both observations the
spectrum was determined using data in the $0.3-10$ keV range.  The
source emission was fit by an absorbed power law during both epochs
using an $N_{H}$\ column density of $6.6 \times 10^{20}$\ cm$^{-2}$\
from Galactic HI \citep{Kalberla2005}.  We measured an X-ray spectral
index of $1.4 \pm 0.1$ during flare B and $1.5 \pm 0.1$\ during the
quiescent state.  The flux in the $0.3-10$ keV range during flare B is
($3.0\pm0.2) \times10^{-12}$ ergs cm$^{-2}$ sec$^{-1}$ and during the
quiescent state the flux is ($3.4\pm0.6) \times10^{-12}$ ergs cm$^{-2}$
sec$^{-1}$.  Therefore the variation observed between the two states is insignificant.

\section{SMARTS Observations}

The Small and Moderate Aperture Research Telescope System (SMARTS)
carries out regular photometric monitoring of LAT sources
of interest at optical and IR frequencies \citep{Bonning2012}\footnote{http://www.astro.yale.edu/smarts/glast/home.php}.
PKS\,2142$-$75 was regularly observed by the ANDICAM instrument
mounted on the SMARTS 1.3\,m telescope from 2010 April 10 (MJD 55296)
to 2012 September 15 (MJD 56185).  The results of the SMARTS
monitoring of this source are displayed in Fig~\ref{gammalightcurve}.
ANDICAM is able to take optical and infrared data simultaneously
with dichroic filters.  A moveable mirror allows dithering
in the IR while the optical exposure is in progress.  The details of the
data reduction are described in \cite{Bonning2012}.  The magnitudes
are corrected for Galactic extinction using the extinction parameters
derived from the recalibration \citep{Schlafly2011} of the
infrared-based dust maps reported in \cite{Schlegel1998}.  Magnitudes
are converted into fluxes using the photometric zero points found in
\cite{Frogel1978}, \cite{Bessell1998} and \cite{Elias1982}.
Simultaneous photometric data points at $B$, $R$, $J$ and $K$ bands
observed on 2010 April 13 (MJD 55299, flare A), 2011 August 12 (MJD
55785, flare B) and 2011 December 28 (MJD 55923, quiescent) are used
in conjunction with our scheduled multi-wavelength observing campaigns
for our SED fits.

\section{Near Infrared Observations with REM}
The Rapid Eye Mount \citep[REM;][]{Molinari2010} telescope is a 60\,cm
telescope observing in the infrared and optical wavebands. It is designed
to react quickly to changes in the sky.  It is located on the La Silla
premises of the ESO Chilean Observatory making it ideal for observing
highly variable southern sources (eg. \cite{Nesci2013}).  Observations of PKS\,2142$-$75, with an exposure time of
150\,s, were made on 2011 August 10 (MJD 55923, flare B) shortly after
the flare in $\gamma$-ray emission observed by LAT.  To
calibrate the data, 10 reference stars from the 2MASS catalog \citep{Skrutskie2006} 
were used with $J$ magnitudes ranging from 11.10 to 14.9.  The linear fit
between instrumental and nominal magnitudes was always good, with a
slope of nearly 1, and a dispersion of about 0.07 magnitude. Two NIR
flux measurements of PKS\,2142$-$75 were made, resulting in magnitudes
of 15.05 $\pm$\ 0.16 in the $J$\ band and 14.92 $\pm$ 0.31 in the $H$\
band. The errors reported are computed by IRAF/apphot and they are
based on the Poisson statistics of the net counts and the detector
noise: this method generally underestimates the actual error in the
source's magnitude. The source was rather faint for REM, so that the
total net counts were rather low and the strongly structured
background light leads to an additional error of 25--30\% (0.3
magnitude) in the flux measurement. Our total error for the source
flux is estimated to be 50\% at both frequencies.

\section{WISE Observations}
The WISE \citep{Wright2010} is
a space based all sky infrared mapping mission.  The cryogen cooled
photometric survey occurred between 2009 January to 2010 October at
3.4--22 $\mu$m.  WISE has a 40\,cm diameter mirror which sits inside a
solid hydrogen cooled cryostat.  WISE only made one observation of our
target source, on 2010 April 13 (MJD 55299).  By a fortunate
coincidence, this observation occurred during a $\gamma$-ray flare of
this object (flare A). These WISE data, at bands $W1, W2, W3, $ and $W4$, were obtained from the All-Sky 
Source Catalogue\footnote{\url{http://irsa.ipac.caltech.edu/cgi-bin/Gator/nph-scan?mission=irsa&submit=Select&projshort=WISE}}. 

%SMARTS and of course LAT
%data were available during this flare although there were no X-ray
%data available for this time period.  

\section{Radio Observations}

The radio data reported here were obtained from two different
monitoring programs, one using the Australia Telescope Compact Array
(ATCA) and the other using the Ceduna radio telescope.  As part of the
TANAMI (Tracking Active Galactic Nuclei with Austral Milliarcsecond
Interferometry) program \citep{Ojha2010}, ATCA has been monitoring
TANAMI sources at several radio frequencies
\citep{Stevens2012}.  ATCA data were reduced with the Miriad software
package, following standard practices for continuum data. Flux
densities were
calibrated against PKS\,1934$-$638 for frequencies below 25\,GHz, and
against Uranus or PKS\,1934$-$638 for higher frequencies. One flux
density measurement was made for each 2\,GHz band, and represents the
flux at the center frequency, assuming that the spectral index across
the band does not vary.  PKS\,2142$-$75 was monitored roughly every
three weeks by the ATCA since flare B.  The results of the monitoring
campaign are displayed in Fig.~\ref{radiofluxes}. Note the rise in the
flux density at 38 and 40\,GHz after flare B.  For the SED fits,
observations on 2011 August 30 (MJD 55803) and 2012 January 15 (MJD
55941) at frequencies of 5.5, 17, 19, 38, and 40\,GHz were used for flare
B and the quiescent period respectively.

%The relative timing and possible correlation in emission from different wavebands is the subject of future work.

The University of Tasmania operates a 30\,m radio telescope at Ceduna
in South Australia \citep[e.g.][]{McCulloch2005}.  This telescope has
been monitoring TANAMI sources at 6.7\,GHz typically once every two
weeks.  The data are collected in four scan blocks, two scans in right
ascension and two in declination. A block is rejected if any of its
scans fail a Gaussian fit. The data are corrected for gain-elevation
effects, and a pointing correction is applied. PKS\,1934$-$638 is used
as the primary flux density calibrator.  Variation in system
temperature due to ambient temperature changes over the course of 24
hours necessitates daily averaging of the data.  The data used in the
construction of the SEDs for flare B and the quiescent period were
obtained on 2011 July 30 (MJD 55772) and 2011 December 15 (MJD 55910).
No significant variation is observed at 6.7\,GHz.  Unfortunately we do
not have radio data during flare A.

\section{Results and Discussion}\label{sed}
%Intro????%
%Models attempt to explain the characteristic SED of blazars and the rapid variability observed in 
%their $\gamma$-ray emission. Further they can calculate (or at least constrain) the physical parameters of the emitting 
%plasma and its turbulence and typical shock speeds. Though it is clear that blazar GeV emission originates in their jets, the physical conditions of %the emitting region are not well understood. To understand the microphysical processes that give rise to $\gamma$-ray emission from 
%blazars the particle distributions and acceleration mechanisms are critical parameters in the modeling process. 
%The variability in GeV emission is useful in constraining acceleration and cooling processes. Further, if all blazars 
%are basically the same kind of objects, the differences in their SEDs might result from processes in the jet i.e. 
%variations in the acceleration mechanism and/or in the dominant particle species are possible. 
%%%%%%%%%%%%% 

%The optical through UV emission is thought to be produced in the accretion disk due to the lack of variability between the three time ranges.  If the %UV optical emission were produce via non thermal synchrotron processes we would expect to see a significant change in the spectrum at those %frequencies in order for the source's broadband emission to be consistent with a leptonic model. Also   The contribution of the synchrotron self %Compton (SSC) effect is very small and plays almost no role. External Compton (EC) processes are the primary source of high energy emission %for this source.  

\begin{figure}
\epsscale{.99}
\plotone{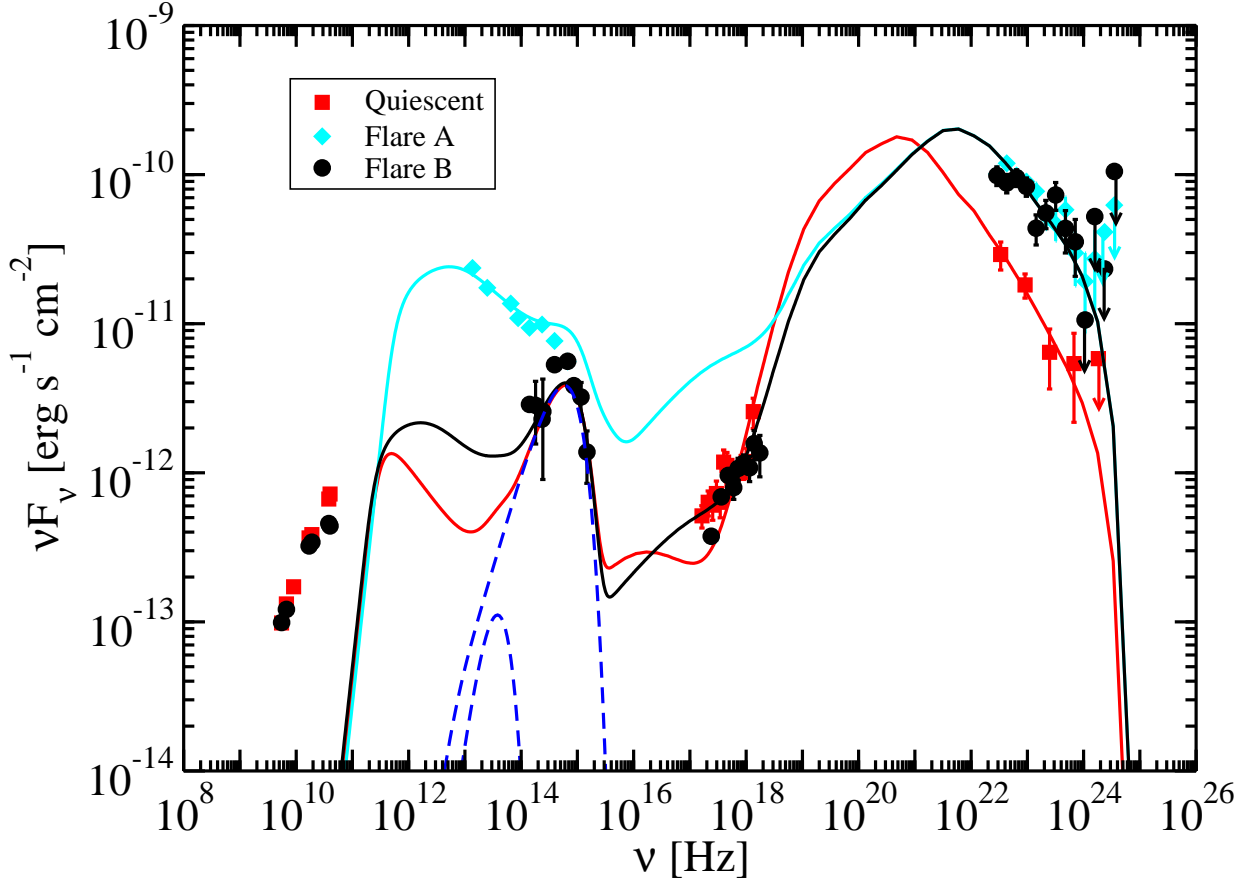}
\caption{Quasi simultaneous spectral energy distributions of PKS\,2142$-$75
measured during active $\gamma$-ray periods in  2010 April (flare A), 2011 August  (flare B) and a quiescent
$\gamma$-ray period in 2011 December.  The data are fit with a
leptonic model of blazar emission. The low energy component is
primarily modeled by synchrotron emission and thermal emission from
the accretion disk. The smaller dashed blue curve represents emission from the dust torus and the larger dashed blue curve is accretion disk emission.  The high energy component is explained by
inverse Compton scattering of dust torus photons by electrons within
the jet.  
\label{SEDfits}}
\end{figure}

The results from each of our three data sets are plotted on the SEDs
shown in Figure~\ref{SEDfits}.  A separate one zone leptonic model of
blazar emission \citep{Dermer2009} was fit for each distinct
$\gamma$-ray state, and the values of the model parameters are
presented in Table~\ref{fitparams_table}.  This model assumes that the
lower-frequency bump is non-thermal synchrotron radiation emitted by
electrons that are isotropically oriented in the co-moving frame of
the jet, in a tangled, randomly oriented magnetic field in the jet
plasma.  Radio emission from the core is thought
to be a superposition of multiple jet components \citep{Konigl1981}
and is therefore treated as an upper limit for the purposes of our
modeling.  We consider SSC and EC processes when modeling the SED of
PKS\,2142$-$75, and we assume that EC scattering is the primary method by
which $\gamma$-rays are produced.  For PKS\,2142$-$75, the size of the emission region is constrained by an upper limit on the variability timescale of 1 day derived from the $\gamma$-ray light curve displayed in Figure \ref{daybin}.

Correlations between $\gamma$-ray and optical flares with radio light
curves and the rotation of optical polarization angles observed in
other LSP blazars \citep{Marscher2010} indicate that the $\gamma$-ray
emitting region may be located outside the broad line region where the dust
torus would be the dominant source of seed photons for external
Compton scattering.  Therefore, for PKS\,2142$-$75 we assumed that
inverse Compton scattering of dust torus photons is the primary method
by which high-energy photons are produced.  The dust torus was assumed
to be a one-dimensional annulus oriented orthogonal to the jet
directions.  Its parameters were chosen to be consistent with the
sublimation radius \citep{Nenkova2008}.

The lack of correlated variability between the $\gamma$-ray flux and
the optical/UV flux in flare B and the quiescent state as well as the
shape of the optical components of the SED during those two states is
indicative of the accretion disk playing a role at those frequencies.
Also a historical optical spectrum, which was used for the
determination of the redshift \citep{Jauncey1978} of this source,
shows broad line emission.  We assume that the accretion disk is an
optically thick blackbody.  The accretion disk parameters and black
hole mass \citep{Shakura1973} were chosen to be consistent with the
optical data in the flaring and quiescent states.  The disk's
luminosity is $L_{\mathrm{disk}}$=0.047 $L_{\mathrm{Eddington}}$ which
is a fairly typical value for FSRQs \citep{Ghisellini2011}.

Flare A does exhibit correlated variability at optical wavelengths.
For flare A the intensity of the magnetic field ($B$) in the model of 
the emitting region needs to 
be changed due to the more dramatic response from the synchrotron
component of the spectrum.  The IR fluxes measured by the SMARTS
team are almost an order of magnitude greater than they were in flare
B.  In order to decrease the ratio of inverse Compton radiation to
synchrotron radiation the magnetic field had to be increased from 0.3
G (its modeled value during flare B and the quiescent period) to 1.0 G.  The
magnetic jet power is an order of magnitude higher than it is during
the quiescent state and flare B.  This increase in the magnetic field
allows the synchrotron flux to increase without having an effect on
the inverse Compton spectrum.  For flare A, our model predicts a
significant X-ray flare from SSC emission.  Unfortunately, no X-ray
observations are available for that time.  If no X-ray flare was
observed, or if a smaller X-ray flare than predicted was observed,
that would argue for a larger $R_b^\prime$ to reduce the SSC emission
observed.  The SEDs could be fit with a longer variability timescale, making the emission region less compact. This would decrease the SSC component, so that other parameters would need to be changed to fit the SEDs.  For example, a combination of lower Doppler factor and higher dust luminosity should also allow a good fit in this scenario.  This should also decrease the jet power in electrons, bringing the model closer to equipartition between electron and magnetic field energy density.  However, this would contradict the short variability timescale found from the light curve (see Figure \ref{daybin} and Section 2 above).

%\begin{figure}[h]
%\includegraphics[width=80mm]{pks2142-75_SED.png}
%\caption{Quasi simultaneous SED of PKS\,2142$-$75 while in a $\gamma$-ray active state.} 
%\end{figure}

%\section{Results and Disscussion}\label{sed}

The electron distribution within the emitting region is described as a
broken power law with indices $p_1$ below and $p_2$ above a break at
$\gamma^\prime_{brk}$ and a low and high energy cutoff at a minimum
and maximum Lorentz factor ($\gamma^\prime_{min}$ and
$\gamma^\prime_{max}$, respectively).  For our source, $p_1$ is not
well constrained and $p_2$ is constrained by the $\gamma$-ray
spectrum.  During flares A and B $\gamma^\prime_{brk}$ increases from
100 (in the quiescent state) to 300, and the spectrum above
$\gamma^\prime_{brk}$ hardens, with $p_2$ decreasing from 4.0 during
quiescence to 3.7 during the active states.  Somewhat
counterintuitively, the jet power in electrons ($P_{j,e}$) decreases
during the active states.  This is because $\gamma_{brk}$ is lower
during the quiescent state, and most of the power is below
$\gamma_{brk}$.  The jet power in Poynting flux ($P_{j,B}$) is the
same in the quiescent and flare B state, since the magnetic field,
$B$, does not change between them; however, it increases by more than
a factor of 10 in the flare A state, due to the increase in $B$.

For flare B we do not increase the magnetic field compared to the
quiescent state.  Only changes to the electron distribution function
are needed.  This is similar to results for PKS\,0537$-$441
\citep{D'Ammando2013}, where it was also found that only changes in the
electron distribution were needed to describe three states of that
source. That is because for the quiescent and flare B states of PKS\,2142$-$75, the
optical emission seems to be dominated by the accretion disk, and we
have modeled it as such.  This makes the synchrotron emission much
less constrained.  The results presented here are, therefore, much
weaker than those of \cite{D'Ammando2013}.  

During all three states the magnetic and electron jet power are far
from equipartition.  This is different from other AGN that have been
modeled in a similar manner
\citep{D'Ammando2012B, D'Ammando2012, D'Ammando2013}.  The jet power in
electrons is much greater than the jet power in magnetic field during
all three states.  Jet powers are calculated assuming that the jet is
two sided \citep{Finke2008}.

%The rise in the high frequency radio emission after flare B might
%indicate the ejection of a new parsec scale radio component, similar to
%those detected by VLBI observations in other flaring sources
%\citep{Marscher2010}.  VLBI observations of this source were scheduled
%after flare B.  Once enough epochs are observed we will be able to
%track the motion and determine the speed of the new component.  When 
%a reliable jet speed
%is obtained we should be able to trace the component back and see if
%the $\gamma$-ray flare was coincident with the new component passing
%through the core.  This would be an important experimental result, possibly 
%indicating that the physical processes that caused the
%$\gamma$-ray emission are related to the VLBI core.

\section{Conclusions}\label{conclusions}

During flare A, PKS\,2142$-$75 was observed with
SMARTS, WISE and {\em Fermi}, which together provide strong constraints on the
synchrotron component of the SED during that flare.  During flare B,
PKS\,2142$-$75 was observed with the ATCA, the Ceduna radio
telescope, REM, SMARTS, {\em Swift} and {\em Fermi}. To provide a
baseline, another multi-wavelength observing campaign was mounted in
2011 December/2012 January when it was in a quiescent state.  By measuring
the brightness simultaneously at many wavelengths we have
gained insight on some of the intrinsic parameters of this highly
variable source.

We have shown that flare B can be explained by changing only the
electron distribution relative to the quiescent state, while flare A
requires changing an additional parameter, namely the magnetic field
strength, $B$.  The two flares in PKS\,2142$-$75 have very different
behavior in the synchrotron component of the spectrum while showing
similar behavior in the $\gamma$-ray component of the spectrum.
Clearly, the same source can exhibit different behaviors for different
flares.  Referring to FSRQs in general, a picture where there are two
types of flares seems to be emerging.  One type of flare, like flare B
in PKS\,2142$-$75 and the flares in PKS\,0537$-$441
\citep{D'Ammando2013}, involves only a change in the electron
distribution from the quiescent state.  Another, like flare A in
PKS\,2142$-$75 and the flares in PKS\,0208$-$512
\citep{Chatterjee2013}, seem to require changes in other parameters,
likely the magnetic field strength and perhaps also the size of
the emitting region.

We have modeled the dominant physical process driving the
luminosity of this source at different wavelengths.  The optical and
ultraviolet emission is primarily caused by thermal accretion disk
emission during flare B and the quiescent state as evidenced by the
shape of the optical SED, and the lack of correlated variability at
optical and UV frequencies during these two different $\gamma$-ray
states.  Our model suggests the jet power in electrons is much greater
than the jet power in the magnetic field.  We have also seen that the
synchrotron component can play a major role at optical wavelengths as
is evidenced by flare A.  We assumed that the primary method by which
$\gamma$-rays are produced in PKS\,2142$-$75 is by inverse Compton
scattering of dust torus photons.

It should be noted that a number of the parameters in these three SED models are
rather poorly constrained and observations are in progress or planned
to refine the parameters and resolve some of the degeneracies in the
one-zone leptonic model used in this work.  As part of the TANAMI program \citep{Ojha2010}, we are
currently gathering VLBI data on the source to potentially resolve the jet
structure and track the kinematics of components within the jet. This
may allow us to put additional constraints on the Doppler factor
and resolve the degeneracy between Doppler factor and the magnetic
field. Also, observations in the sub-millimeter regime (especially at
the high frequency end of ALMA's observing band $\sim 10^{12}$\,Hz)
will put important constraints on the emission mechanisms in this
object.

% Further constraints on the electron distribution can be placed by filling in more of the low energy synchrotron spectrum: data from the ALMA (REF) or APEX (REF) telescopes would be particularly helpful.

%Collecting and analyzing Very Long Baseline Interferometry (VLBI) data produced by the Long Baseline Array (LBA) located in the southern hemisphere will allow us to resolve the jet structure of this source. Multiple epochs of VLBI data will also allow us to track the speed of individual jet components which will help us put constraints on the Doppler factor and magnetic field since there is a degeneracy between these two parameters in this model. Also observations in the sub millimeter regime (especially at the high frequency end of ALMA's observing band $\sim 10^{12}$\,Hz) and far infra red will put further constraints on the emission mechanisms in this object. If the spectral shape at sub millimeter and far infrared frequencies indicate that there is not a distinct synchrotron and accretion disk peak in the emission we would be forced to consider the possibility that the optical/UV emission from this source is caused by synchrotron emission as well. If this is the case then a leptonic model which calls for an increase in the jet power in electrons to account for gamma ray variability would not be the best way to describe the emission in this source. 

\acknowledgments
\section{Acknowledgements}

We thank the {\em Swift} team for scheduling our Target of Opportunity requests. This research was funded in part by NASA through {\em Fermi} Guest Investigator grants NNH09ZDA001N and NNH10ZDA001N. This research was supported by an appointment to the NASA Postdoctoral Program at the Goddard Space Flight Center, administered by 
Oak Ridge Associated Universities through a contract with NASA. 
This publication makes use of data products from the {\em Wide-field Infrared Survey Explorer}, which is a joint project of the University of California, Los Angeles, and the Jet Propulsion Laboratory/California Institute of Technology, funded by the National Aeronautics and Space Administration.
The Australia Telescope Compact Array is part of the Australia Telescope National Facility which is funded by the Commonwealth of Australia for operation as a National Facility managed by CSIRO.
This research has made use of data from the NASA/IPAC
Extragalactic Database (NED), operated by the Jet Propulsion Laboratory,
California Institute of Technology, under contract with the National
Aeronautics and Space Administration; and the SIMBAD database (operated
at CDS, Strasbourg, France). T his research has made use of NASA's
Astrophysics Data System. This research has made use of the United States
Naval Observatory (USNO) Radio Reference Frame Image Database (RRFID).  This paper has made use of up-to-date SMARTS optical/near-infrared light curves that are available at \url{www.astro.yale.edu/smarts/glast/home.php}.

The {\em Fermi} LAT Collaboration acknowledges generous ongoing support
from a number of agencies and institutes that have supported both the
development and the operation of the LAT as well as scientific data analysis.
These include the National Aeronautics and Space Administration and the
Department of Energy in the United States, the Commissariat \`a l'Energie Atomique
and the Centre National de la Recherche Scientifique / Institut National de Physique
Nucl\'eaire et de Physique des Particules in France, the Agenzia Spaziale Italiana
and the Istituto Nazionale di Fisica Nucleare in Italy, the Ministry of Education,
Culture, Sports, Science and Technology (MEXT), High Energy Accelerator Research
Organization (KEK) and Japan Aerospace Exploration Agency (JAXA) in Japan, and
the K.~A.~Wallenberg Foundation, the Swedish Research Council and the
Swedish National Space Board in Sweden.

Additional support for science analysis during the operations phase is gratefully
acknowledged from the Istituto Nazionale di Astrofisica in Italy and the Centre National d'\'Etudes Spatiales in France.

{\it Facilities:} \facility{ATCA}, \facility{Ceduna Radio Observatory}, \facility{Fermi}, \facility{REM}, \facility{Swift}, \facility{SMARTS}.

%\appendix

%\section{Appendix material}

\bibliographystyle{apj}
\bibliography{aa_abbrv,mnemonic,tanami}

\clearpage

\begin{deluxetable}{lccccccc}
\rotate
\tabletypesize{\scriptsize}
\tablecaption{
Model fit parameters
}
\tablewidth{0pt}
\tablehead{
\colhead{Parameter} & 
\colhead{Symbol} & 
\colhead{Quiescent} & 
\colhead{Flare A} & 
\colhead{Flare B} 
}
\startdata
\hline
Bulk Lorentz Factor                        & $\Gamma$	           & 30                 & 30                 & 30 	          \\
Doppler Factor                             & $\delta_D$            & 30                 & 30                 & 30                 \\
Magnetic Field [G]                         & $B$                   & 0.3                & 1.0                & 0.3                \\
Variability Timescale [s]                  & $t_v$                 & $6.0\times10^4$    & $6.0\times10^4$    & $6.0\times10^4$    \\
Comoving Blob radius [cm]                  & $R_b^\prime$          & $2.5\times10^{16}$ & $2.5\times10^{16}$ & $2.5\times10^{16}$ \\
\hline								   													     
Low-Energy Electron Spectral Index         & $p_1$                 & 2.2                & 2.2                & 2.2                \\
High-Energy Electron Spectral Index        & $p_2$                 & 4.0                & 3.7                & 3.7                \\
Minimum Electron Lorentz Factor            & $\gamma^\prime_{min}$ & $11$              & $11$              & $11$              \\
Break Electron Lorentz Factor              & $\gamma^\prime_{brk}$ & $1.0\times10^2$    & $3.0\times10^2$    & $3.0\times10^2$    \\
Maximum Electron Lorentz Factor            & $\gamma^\prime_{max}$ & $7.0\times10^3$    & $7.0\times10^3$    & $7.0\times10^3$    \\
\hline								   													     
Black hole mass [$M_\odot$]                & $M_{BH}$              &  $5\times10^9$     & $5\times10^9$      &  $5\times10^9$     \\
Disk luminosity [erg s$^{-1}$]             & $L_{disk}$            & $2.9\times10^{46}$ & $2.9\times10^{46}$ & $2.9\times10^{46}$ \\
Accretion efficiency                       & $\eta$                & 1/12               & 1/12               & 1/12               \\
Jet Height [cm]                            & $r$                   & $3.1\times10^{18}$ & $3.1\times10^{18}$ & $3.1\times10^{18}$ \\
\hline								   													     
Dust temperature                           & $T_{dust}$            & $1.0\times10^{3}$  & $1.0\times10^{3}$  & $1.0\times10^{3}$  \\
Dust luminosity [erg s$^{-1}$]             & $L_{dust}$            & $1.1\times10^{45}$ & $1.1\times10^{45}$ & $1.1\times10^{45}$ \\
Dust Radius [cm]                           & $R_{dust}$            & $9.7\times10^{18}$ & $9.7\times10^{18}$ & $9.7\times10^{18}$ \\
\hline								   													     
Jet Power in Magnetic Field [erg s$^{-1}$] & $P_{j,B}$             & $3.9\times10^{44}$ & $4.3\times10^{45}$ & $3.9\times10^{44}$ \\
Jet Power in Electrons [erg s$^{-1}$]      & $P_{j,e}$             & $2.1\times10^{46}$ & $1.2\times10^{46}$ & $1.2\times10^{46}$ \\
\enddata
\label{fitparams_table}
\end{deluxetable}

\clearpage

\clearpage

\end{document}